\documentclass[pre,aps,twocolumn,showpacs]{revtex4-2}
\usepackage{graphicx}
\usepackage{dcolumn}
\usepackage{bm}
\usepackage{color}
\usepackage{url}
\usepackage{amssymb}
\usepackage{amsmath}
\usepackage [latin1]{inputenc}
\usepackage{float}
\usepackage{mathtools}
\usepackage{tikz}
\newcommand{\tikzcircle}[2][red,fill=red]{\tikz[baseline=-0.5ex]\draw[#1,radius=#2] (0,0) circle ;}

\begin{document}

\title{How capture affects polymer translocation in a solitary nanopore}

\author{Swarnadeep Seth}
\author{Aniket Bhattacharya}

\altaffiliation[]
{Author to whom the correspondence should be addressed}
{}
\email{Aniket.Bhattacharya@ucf.edu}
\affiliation{$^1$Department of Physics, University of Central Florida, Orlando, Florida 32816-2385, USA}
\date{\today}

\begin{abstract}
DNA capture with high fidelity is an essential part of nanopore translocation. We report several important aspects of the capture process and subsequent translocation of a model DNA polymer through a solid-state nanopore in presence of an extended electric field using the Brownian dynamics simulation that enables us to record statistics of the conformations at every stage of the translocation process. By releasing the equilibrated DNAs from different equipotentials, we observe that the capture time distribution depends on the initial starting point and follows a Poisson process. The field gradient elongates the DNA on its way towards the nanopore and favors a successful translocation even after multiple failed threading attempts.
Even in the limit of an extremely narrow pore, a fully flexible chain has a finite probability of hairpin-loop capture while this probability decreases for a stiffer chain and promotes single file translocation. Our {\em in silico} studies identify and differentiate characteristic distributions of the mean first passage time due to single file translocation from those due to translocation of different types of folds and provide direct evidences of the interpretation of the experimentally observed folds [M. Gershow {\em et al.}, Nat. Nanotech. {\bf 2}, 775 (2007) \& M.  Mihovilovic {\em et al.} Phys. Rev. Letts. {\bf  110}, 028102 (2013)] in a solitary nanopore. Finally, we show a new finding, - that a charged tag attached at the $5^{\prime}$ end of the DNA enhances both the multi-scan rate as well as the uni-directional translocation ($5^{\prime} \rightarrow 3^{\prime}$)
probability that would benefit the genomic barcoding and sequencing experiments.
\end{abstract}
\maketitle
\section{Introduction}
Efficient capture of a dsDNA  is the first step for its subsequent translocation through the nanopore that is central for sequencing, genome mapping, and other multiplexed nanopore sensing applications
~\cite{Deamer2016-Review,Metzler2014_Review,Wanunu2012_Review,Keyser2011_review,Muthukumar2011_Review,Bashir2011_Review}. A voltage is applied  across the nanochannel to electrophoretically drive a dsDNA or other biopolymer through the nanopore or a nanochannel. The characteristic current blockade data is then deconvoluted to reconstruct the translocated species \cite{Branton2001}-\cite{Denniston2020}.
The capture rate in a nanopore device is a function of several adjustable parameters, the pH of the electrolyte, the diameter and shape of the nanopore, the strength of the external electric field, and the length of the biomolecule and its effective charge. Evidently, a fundamental understanding of the dependencies of the capture rate on these factors is essential to improve efficiency and quality of a nanopore device. A large majority of these items have been addressed experimentally~\cite{Dekker2005,Golovchenko2007,Wanunu2009,Stein2013,Jeon-Muthukumar_2014}. The theoretical studies have either concurred or validated some of these dependencies~\cite{Muthukumar2010,Grosberg2010,Grosberg2013}.  Parallelly, simulation studies~\cite{Hann2016,Aksimentiev2020,Slater2019,Slater_PCCP2020,Slater_JCP2020} have revealed more detailed picture of the capture process and helped towards the construction of a unifying theory of capture and translocation.
\par
Nanopore translocation consists of three distinct processes. Diffusion of a polymer towards the pore which progressively acquires characteristics of a drift as it approaches the pore, its eventual capture, often after multiple attempts, and finally, the successful threading through the nanopore which we call translocation.
A large fraction of the earlier simulation studies were done assuming the polymer is already captured at the pore entrance and that a localized field exists only inside the pore that drives the biopolymer from the {\em cis} to the {\em trans} side. One of the major focus of these theoretical and computational studies were directed to find how the mean first passage time $\langle \tau \rangle$ depends on the chain length assuming a power law dependence $\langle \tau \rangle \sim N^{\alpha}$, where $\alpha$ is the translocation exponent~\cite{Muthukumar2011_Review,Metzler2014_Review}.
However, the diffusion and the drift of the polymer, and its eventual capture in the pre-translocation phase getting ready to be translocated itself is very rich in physics as has been demonstrated in recent  experimental studies studies~\cite{Golovchenko2007, Wanunu2009,Stein2013,Jeon-Muthukumar_2014}. 
Gershow and Golovchenko studied Kilo-base-pair (kbp) long dsDNA and studied capture-recapture probability as a function of the delay time (which approximately translates to different distances from the pore) of the voltage reversal ~\cite{Golovchenko2007}. Mihovilovic {\em et al.} made a detailed study of how capture of a folded configuration affects its translocation property~\cite{Stein2013}. Wanunu et al. found that for a fixed concentration, the capture rate increases with the contour length up to a critical length beyond which it saturates~\cite{Wanunu2009}. They further observed that a salt gradient increases the capture probability. 
Jeon and Muthukumar conducted a similar study on the dependence of salt-concentration gradient and pore-polymer interaction using a $\alpha$-haemolysin protein pore~\cite{Jeon-Muthukumar_2014}. These experimental studies have motivated more simulation studies on the capture problem. Slater and co-workers used different approaches to define the capture radius, studied the effect of the time dependent field~\cite{Slater2019,Slater_PCCP2020,Slater_JCP2020} on the capture process. Vollmer and deHaan studied the change in the shape of the polymer as it approaches to the pore by the ratio of gyration radii along the longitudinal and transverse direction~\cite{Hann2016}. 
\par
\begin{figure*}[ht!]
\includegraphics[width=0.98\textwidth]{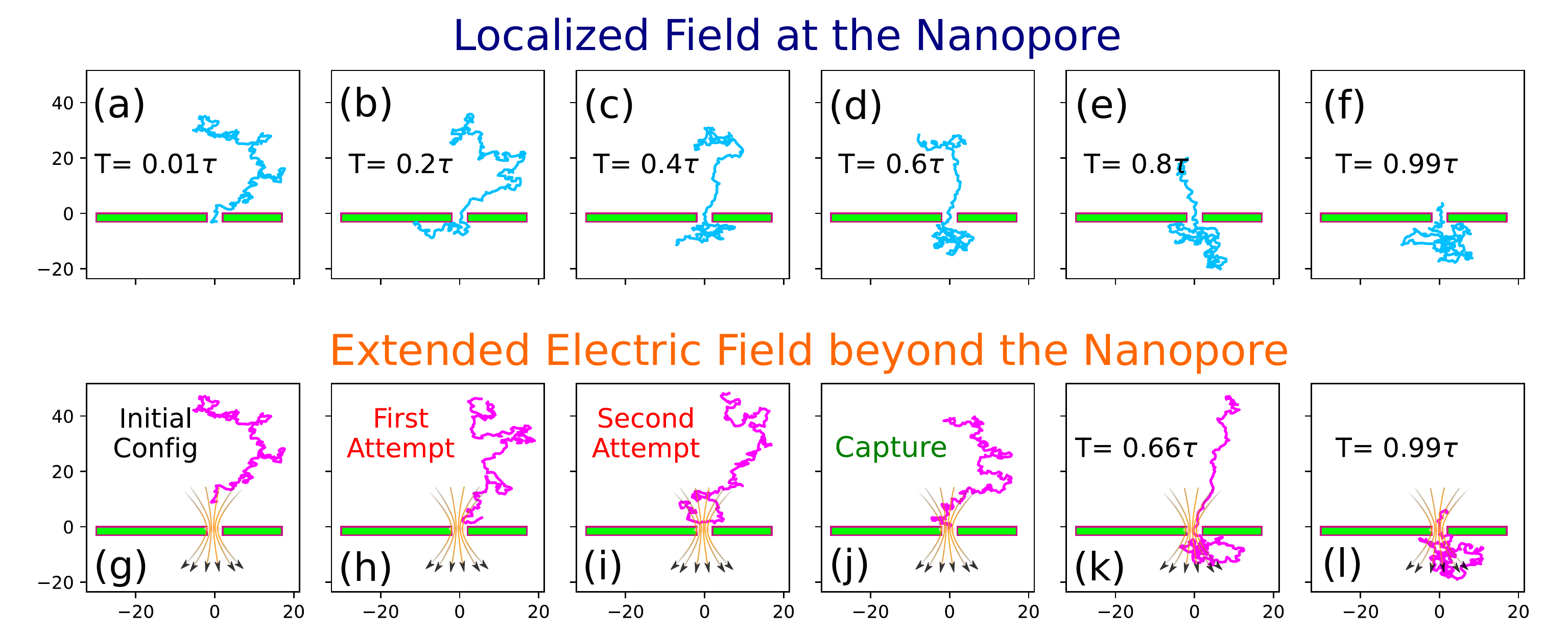} 
\caption{\small  A series of simulations snapshots (a)-(f) in progressive time show the translocation process of the polymer in presence of localized force bias applied in the nanopore thickness. $\tau$ is the total translocation time of the chain. (g)-(l) The same polymer configuration is released from the equipotentials surface $d=4.0$ away from the pore in presence of the electric field which extends beyond the nanopore. During the capture process, the polymer makes a few unsuccessful attempts shown in (h) and (j) before one end threads into the nanopore. (k)-(l) show the translocation of the captured polymer through the nanopore under the electric field gradient.} 
\label{Model}
\end{figure*}

In this article, we report a few new results on further details of the capture process and how the capture affects the translocation properties of a biopolymer. A few results have been influenced by and complement previous studies and a few others which we believe add to our overall understanding of the capture and translocation characteristics. We primarily consider three factors; the extended electric field outside the pore, the initial release point of the polymer, and the DNA persistence length on the capture and translocation probability keeping the length and the field strength constant and present the following key results.
(i) We revisit the effect of the electric field and demonstrate how an extended field influences the capture process as opposed to a field that strictly resides inside the nanopore. The polymer makes multiple attempts in presence of an extended field (Fig.~\ref{Model}), similar to Kramer's barrier crossing problem~\cite{Kramer1940} and increases the capture probability, as recently theoretically discussed by Grossberg~\cite{Grosberg2010,Grosberg2013}. (ii) Following Vollmer {\em et al.} we use the ratio of the gyration ellipses to monitor the shape of the polymer which shows a signature as the diffusive motion acquires characteristics of a drift as the polymer approaches the pore (Fig.~2). (iii) We further show that releasing the polymer close to the pore mouth from stronger equipotentials has higher success rate of capture (Fig.3) and shorter capture time (Fig. 4(a)-(d)) and that (iii) the release distance affects the formation and translocation of hairpin structures (Fig.~4). (iv) We provide further details and validate different modes of folded and unfolded configurations from the reversal of the electric field (Fig.~5). We then demonstrate that the single-file capture can be enhanced drastically by stiffening the polymer (Fig. 6) and the hairpin translocation probability is negligible after achieving a critical persistence length. (v). Finally, we show that an effective way to increase the unidirectional capture rate is to attach charged tags at one end the DNA which breaks the degeneracy promoting a higher rate of unidirectional capture with the increasing stiffness (Fig.7), often required for the nanopore sequencing experiments. Thus the manuscript connects new results with those already published, reveals the details of the experimentally observed modes of translocation, and provides a unifying picture of how the capture occurs and how different captured configurations eventually translocate through the nanopore. 
\par
\section{Coarse-grained model \& Langevein dynamics simulation:} 
Our CG model of a dsDNA consists of 256 beads of diameter of $\sigma$, that mimics a $4 \mu$m long $\lambda$-phage DNA with  $48$ bp resolution associated with a single bead. We use a short range Lennard-Jones (LJ) potential
\begin{eqnarray}
U_{\mathrm{LJ}}(r) &=&4\epsilon \left[{\left(\frac{\sigma}{r}\right)}^{12}-{\left(\frac{\sigma} {r}\right)}^6\right]+\epsilon,
                              \;\mathrm{for~~} r\le 2^{1/6}\sigma; \nonumber\\
                             &=& 0, \;\mathrm{for~~} r >  2^{1/6}\sigma.
\label{LJ}
\end{eqnarray}
to model the excluded volume interaction between two beads separated at a distance $r$ where, $\epsilon$ is the strength of the LJ potential. The connectivity between two neighboring monomers is constructed using the FENE spring potential
\begin{equation}
U_{\mathrm{FENE}}(r_{ij})=-\frac{1}{2}k_FR_0^2\ln\left(1-r_{ij}^2/R_0^2\right).
\label{FENE}
\end{equation}
Here, $r_{ij}=\left | \vec{r}_i - \vec{r}_j \right|$ is the distance between two consecutive monomer beads $i$ and $j=i\pm1$ at $\vec{r}_i$ and $\vec{r}_j$, $k_F$ is the spring constant and $R_0$ is the maximum allowed separation between two connected monomers. 
An angle dependent three body interaction term is introduced between successive bonds which accounts for the chain stiffness $\kappa$
\begin{equation}
U_{\mathrm{bend}}(\theta_i) = \kappa\left(1-\cos \theta_i\right) 
\end{equation}
 and $\theta_i$ is the angle between the bond vectors 
$\vec{b}_{i-1} = \vec{r}_{i}-\vec{r}_{i-1}$ and 
$\vec{b}_{i} = \vec{r}_{i+1}-\vec{r}_{i}$, respectively. 
For a homopolymer chain the persistence length $\ell_p$ in three dimensions (3D) is given by
\begin{equation}
\ell_p/\sigma = \kappa/k_BT.
\label{lp_bulk}
\end{equation}
\par 
A cylindrical nanopore of diameter of $2\sigma$ is constructed by removing out particles from the center of a $2\sigma$ thick wall consists of immobile LJ particles.  
We use the Langevin dynamics simulation following the equations of motion for the i$^{th}$ monomer 
\begin{equation}
m \ddot{\vec{r}}_i = -\nabla (U_\mathrm{LJ} + U_\mathrm{FENE} + U_\mathrm{bend} \\
                                            + U_\mathrm{wall}) -\Gamma \vec{v}_i + \vec{\eta}_i . 
\label{langevin}                                          
\end{equation}
Here $\vec{\eta} _ i (t)$ is the Gaussian white noise with zero mean at temperature $T$, and satisfies the fluctuation-dissipation relation in $d$ physical
dimensions (here $d=3$):
\begin{equation}
< \, \vec{\eta} _ i (t) 
\cdot \vec{\eta} _ j (t') \, > = 2dk_BT \Gamma \, \delta _{ij} \, \delta (t - t ').
\end{equation}
We express length and energy in units of $\sigma$ and $\epsilon$, respectively. The parameters for the FENE potential in Eq.~(\ref{FENE}), $k_F$ and  $R_0$, are set to $k_F = 30 \epsilon/\sigma$ and $R_0 = 1.5\sigma$, respectively. The friction coefficient and the temperature are set to $\Gamma = 0.7\sqrt{m\epsilon/\sigma^2}$ and $k_BT/\epsilon = 1.0$. The numerical integration of Equation~(\ref{langevin}) is implemented using the algorithm introduced by Gunsteren and Berendsen~\cite{Berendsen1982}.
\par
\begin{figure*}[ht!]
\includegraphics[width=0.95\textwidth]{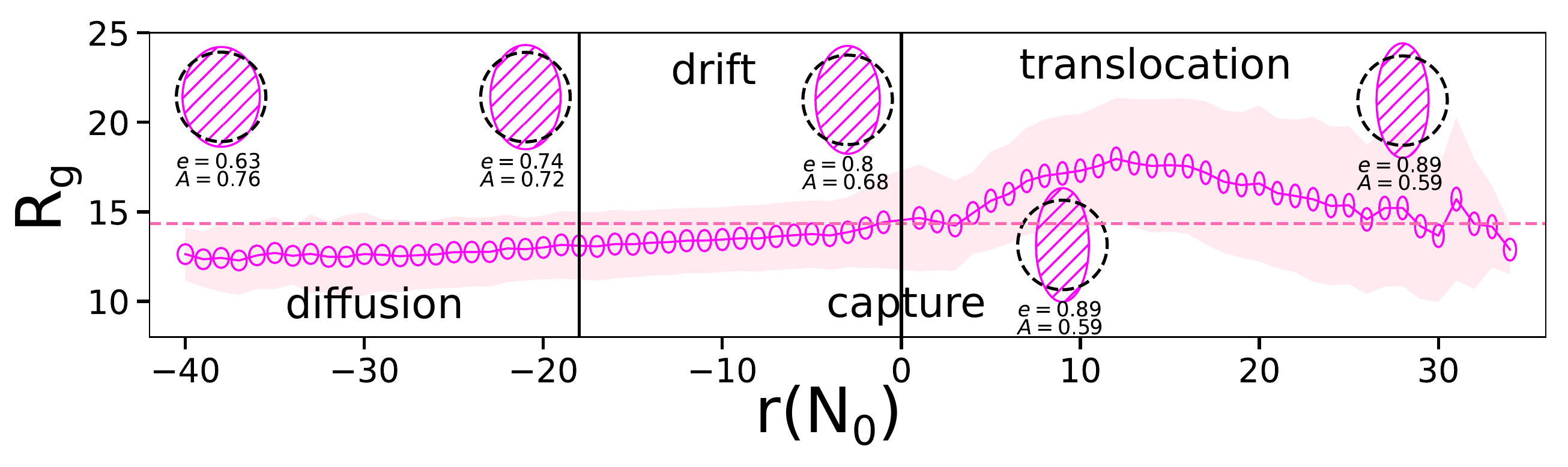} 
 \caption{\small Polymer radius of gyration as a function of first monomer distance $r(N_0)$ from the pore. The effective radius of gyration of the polymer at a distance is represented by an ellipse having the major/minor is representing transverse/longitudinal component respectively. The average $R_g$ is denoted by the magenta dotted line. $R_g$ at different phases (a) diffusion dominated (b) drift dominated and capture (c) translocation and escape are presented in the inset ellipses and compared against a unit circle in black. $e$ and $A$ denote the eccentricity and the area of the ellipses respectively. The drift region is identified from the diffusion dominated region by marking the inflection point of the $R_g$.} 
\label{Rg-Ellipse}
\end{figure*}
\section{Results}
Unless otherwise specified, we choose a polymer of length $N=256$ and each monomer carries a charge $|q_{i}|=1.0$. We release the equilibrated polymers from different equipotentials $\Phi$ referenced using the vertical distances $d/\sigma =4, 8,12,$ and $16$ from the pore (corresponding the equipotential values $\Phi = 0.322, 0.157, 0.101,$ and $0.071$ in MD units respectively, please see Table-I). The release configuration of a polymer is such that the first monomer stays on the equipotential $\Phi$ and rest of the chain conformation remains in the lower equipotential levels away from the pore. The equipotentials are calculated numerically but exactly with appropriate boundary conditions and closely resemble those using the analytic expression by Farahpour et al.~\cite{Farahpour2013}. Far away from the pore, the equipotentials are almost concentric circles but close to the pore become elliptical. We consider fully flexible chain and chains of persistence length corresponding to $\kappa = 3,6,$ and $9$ respectively. The initial locations of the polymer are then chosen by uniform random sampling for a given equipotential. In addition, for certain cases, we carried out simulation for a chain length $N=512$. For the first translocation event, we choose the negative $y$-axis to be the translocation axis (Fig.~\ref{Model}), the captures occur at $y = 0$, and the electric field deep inside the pore is directed along the $-y$ axis (Fig.~\ref{Model}(b)). Both the diameter and width of the nanopore are chosen to be  $2\sigma$ that also allows translocation of the folded configurations. For the multiple scans the electric field is reversed accordingly across the pore. 
\subsection{Translocation in a localized electric field:}
All of our results excepting Fig.~\ref{Model}(a)-(f) are presented for an electric field that extends beyond the pore. To contrast those results we show the results for an equilibrated chain for four Rouse relaxation time with an initial configuration placing a few beads inside the nanopore those experience the localized electric field strictly inside the pore. At the start of the translocation process, the beads inside the nanopore experience a downward pulling force while the tension front~\cite{Sakaue2007} propagates through the chain backbone in the opposite direction of translocation that first uncoils the chain. The chain is quickly sucked into the pore when the tension front hits the last bead~\cite{Adhikari2013}. This entire process is demonstrated in Fig.~\ref{Model} ((a)-(f)). 
\subsection{Capture in an extended electric field:}
However, the process in Fig.~\ref{Model} ((a)-(f)) does not resemble an experimental situation. In an experimental situation the polymer is first released into the solution executes a drift-diffusive motion and gets captured aided by the extended electric field beyond the pore shown in Fig.~\ref{Model} ((g)-(l)). We use the Finite Element method \cite{Fenics2012} to solve the Poisson equation to solve the electric field around the solitary nanopore. Unlike the case of a localized E-field, the extended E-field elongates the polymer polymer along the field vector.  The field
gradient, which is strongest at and near the pore also helps the capture process. Vollmer {\em et al}. have shown that the radius of gyration ellipse deforms differently depending on the polymer size and the P\'{e}clet number~\cite{Hann2016}.  We use this idea, but monitor the average radius of gyration $\langle R_g \rangle$ for the entire processes from capture to translocation that shows the shape of the polymer under the influence of an extended electric field. We characterize this dynamic deformation process by comparing the transverse and longitudinal radius of gyration as the polymer drifts towards the pore. We define the longitudinal and the transverse gyration radii as in Eqn.~\ref{ellipse_long} and \ref{ellipse_perp}, 
\begin{subequations}
\begin{equation}
\langle Rg_{||}\rangle = \sqrt{\langle R_y^2 \rangle} \label{ellipse_long}
\end{equation}
\begin{equation}   
\langle Rg_{\perp}\rangle  = \sqrt{\langle Rg_{x}^2 \rangle + \langle Rg_{z}^2 \rangle } \label{ellipse_perp}
\end{equation}
\begin{equation}
\langle Rg_{||}\rangle / \langle Rg_{\perp}\rangle \label{ellipse}
\end{equation}
\end{subequations}
and construct a radius of gyration ellipse using Eqn.~\ref{ellipse}. Far away from the pore, the polymer remains almost unaffected by the electric field, resembling the equilibrium configuration with $\langle Rg_{||}\rangle \simeq \langle Rg_{\perp}\rangle $. As polymer drifts along the field lines, $\langle Rg_{||}\rangle$ extends over $\langle Rg_{\perp}\rangle $ enhancing the eccentricity of the gyration ellipse as shown in Fig.~\ref{Rg-Ellipse}. The gyration ellipse eccentricity increases steadily until the polymer translocates through the nanopore. Specifically, how the electric field affects the post-translocational conformations are further discussed in Section~E.
\par
\subsection{Polymer drift and capture at the nanopore}
The process of capture requires the polymer to overcome the potential barrier by adjusting its conformational entropy resembling Kramer's escape problem~\cite{Kramer1940,Wanunu2009}. The capture is a non-equilibrium process~\cite{Sakaue2007,Adhikari2013,Muthukumar2018} where field strength dominates over the diffusion near the pore. This  leads to a directed motion of either end of the polymer until the polymer gets captured. During simulation, the equilibrated DNA polymers are released from different equipotentials. To implement that we first determine the potential $\Phi$ at a vertical distance $d$ from the nanopore orifice and locate other points on the same equipotential surface. We use these equipotentials as the starting locations of the DNA polymer as this would be easier to compare and perform in the experiments.
We define the capture probability $P_{cap}(\Phi,t)$ of a polymer at the pore at time $t$ released from an equipotential $\Phi$ at time $t=0$ as 
\begin{equation}
  P_{cap}(\Phi,t)=\frac{N_p(\Phi, t)}{N_p(\Phi,0)},
\end{equation}
where $N_p(\Phi,t)$ is the number of polymers captured at time $t$ at the pore and $N_p(\Phi,0)$ is the number released at the beginning at time $t=0$ from an equipotential $\Phi$. The capture of polymers at the pore from an equipotential are independent events (as they are released sequentially one after another), thus follows a ``memoryless" Poisson process and time between capture events resembles an exponential distribution. This has been measured experimentally also in a single nanopore context~\cite{Branton2002}. Therefore, theoretically the capture probability is the cumulative distribution function of the exponential distribution
\begin{equation}
P_{cap}(\Phi,t) = 1 - \exp \left (-\lambda_{cap} t \right).
\label{P_cap}
\end{equation}
\begin{figure}[ht!]
\includegraphics[width=0.48\textwidth]{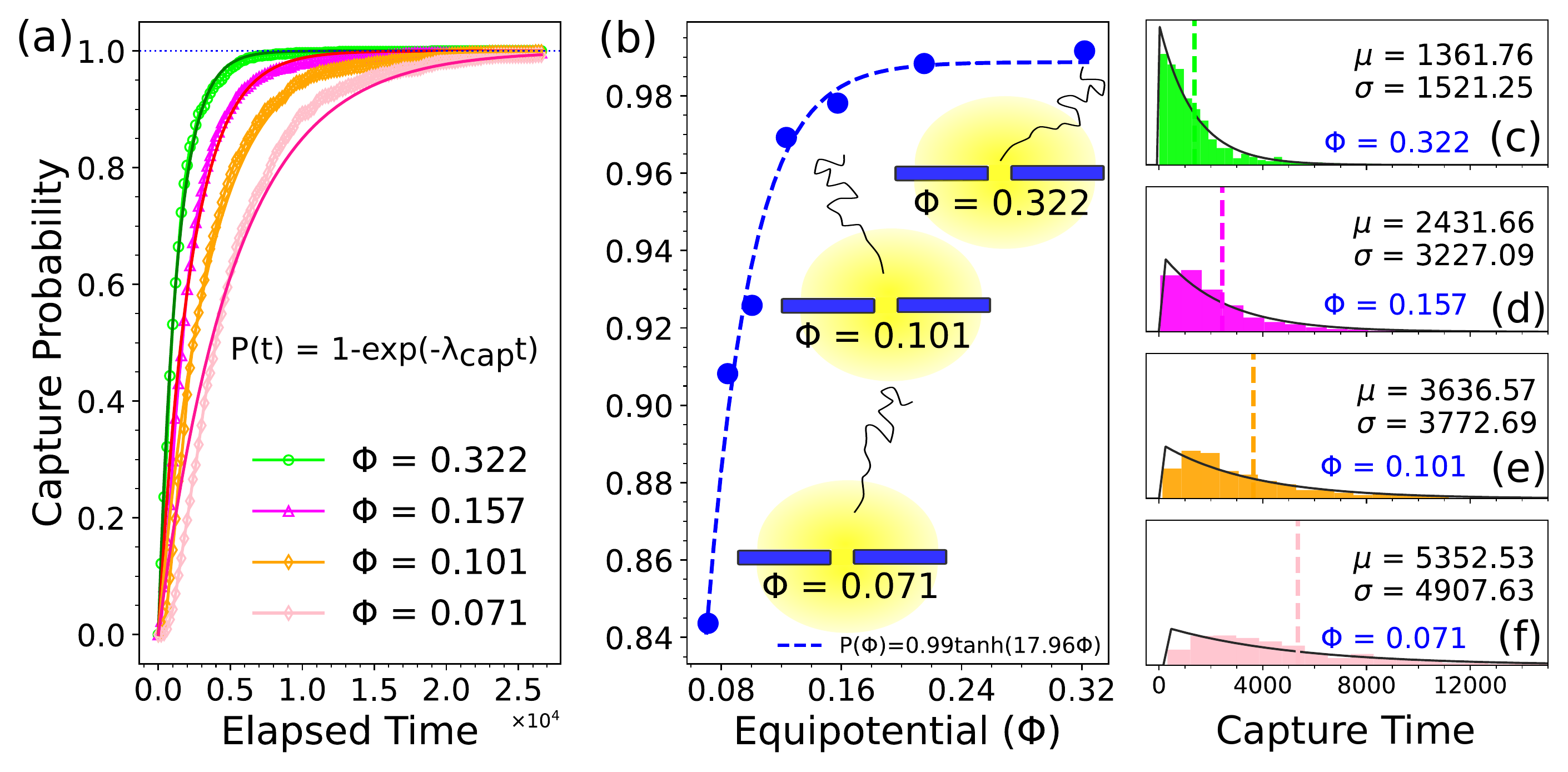} 
\caption{\small (a) Capture probability $P_{cap}(\Phi,t)$ as a function of time for the polymer released from different equipotentials  $\Phi=0.322$  (\tikzcircle[green, fill=none]{2.5pt}), $0.157$ (${\color{magenta}{\bigtriangleup}}$), $0.101$ (${\color{orange}{\Diamond}}$), and $0.071$ (${\color{red}{\Diamond}}$). In each case the solid colored line represents a fit to Eqn.~\ref{P_cap} with $\lambda_{cap}=\frac{1}{\mu}$ shown in Table-I. (b) Dependence of the capture probability $P_{cap}(\Phi,t)$ as a function of the equipotential $\Phi$. The line through the points is described by a function $A \tanh(\beta \Phi)$. The graphs (c)-(f) show the capture time distributions for different equipotentials. In each case, $\mu$ and $\sigma$ corresponds to the distribution average and standard deviation respectively. The black envelops show the exponential fits of the distributions with the averages marked by dashed lines.}
\label{Capture-Prob}
\end{figure}
The cumulative capture probabilities for different equipotentials obtained from BD simulation are shown in Fig~\ref{Capture-Prob} (a). $P_{cap}(\Phi,t)$ increases with time and follows Eqn.~\ref{P_cap}. It is important to note that the values of the prefactor $\tilde{\lambda}_{cap}$ to best fit the Eqn.~\ref{P_cap} in Fig.~\ref{Capture-Prob}(a) are almost the same as the values  $\lambda_{cap}= \frac{1}{\mu}$  obtained from the capture time distributions (Fig.~\ref{Capture-Prob}(c)-(f)) as shown in Table-\ref{capture_table}. 
\begin{figure}[ht!]
\includegraphics[width=0.4\textwidth]{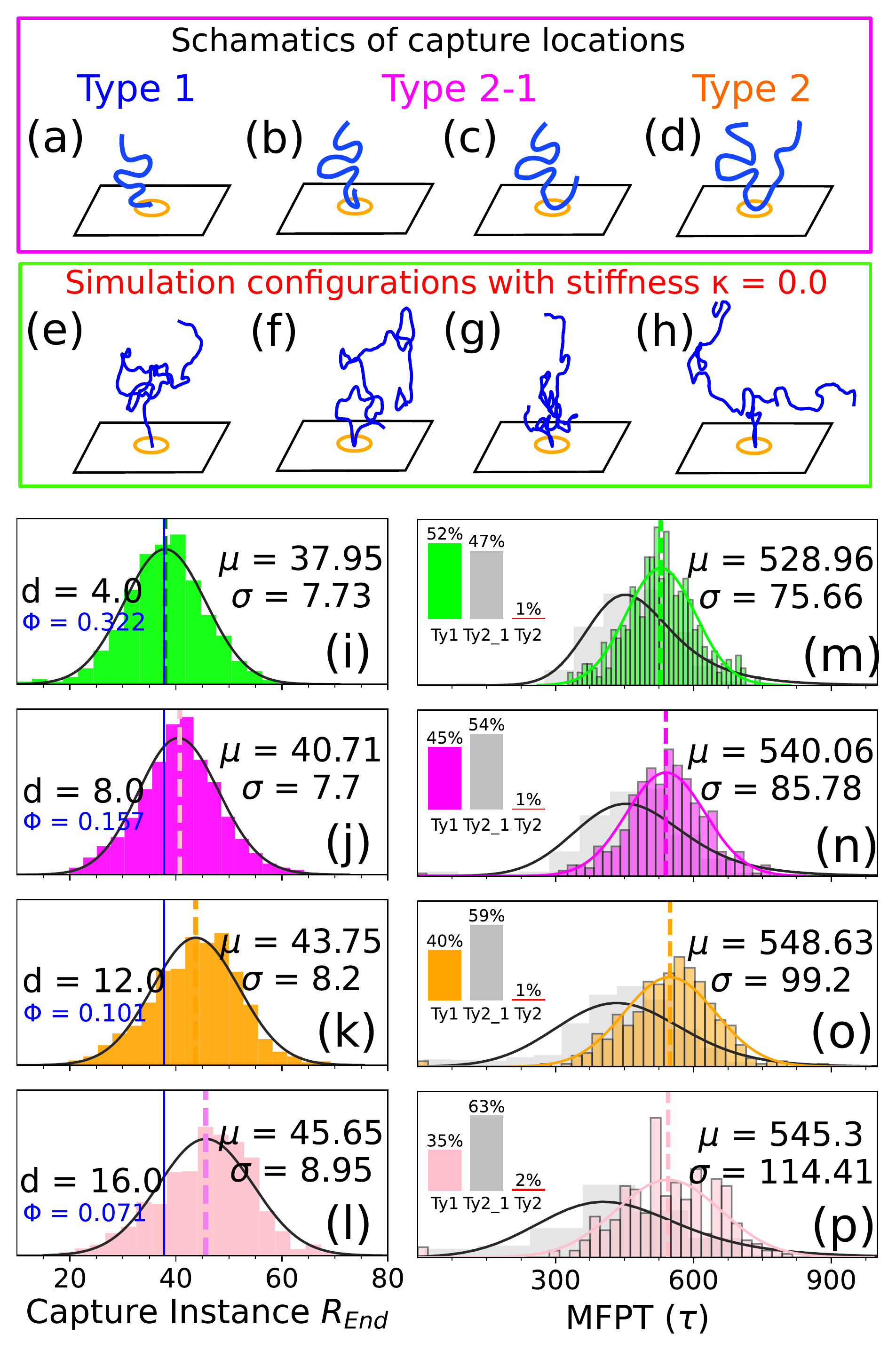}
\caption{\small (a)-(d) show the rendition of the different ways a polymer threads through the nanopore and translocates. The capture location  $x$ denotes the normalized monomer index $m/N$ which first threads into the nanopore; (a) The single file capture events where either of the ends gets captured are denoted as Type1 translocation. (b)-(c) show the Type2-1 events in which polymer is captured at any random location except at the ends and in the symmetric location. (d) Type2 events indicate the symmetrical capture cases. (e)-(h) denote the same using the actual coordinates from BD simulation for a fully flexible polymer. End to end distance $R_{end}$ distributions of a polymer at the moment of capture at the nanopore orifice after being released from four different equipotential distances $d=4.0$, $8.0$, $12.0$, and $16.0$ are shown in (i)-(l) sub-plots. The black envelops denote the exponentially modified Gaussian fits of the distributions. Average and standard deviation of the distributions are marked in $\mu$ and $\sigma$ respectively and corresponding $\langle R_{end} \rangle$ are shown in the colored dashed lines while the blue line represents the average end to end distances of the starting configurations.
After capture, the mean first passage time histograms (m)-(p) follow Gaussian shape with average $\mu$ and standard deviation $\sigma$ are obtained from single file translocation (Type 1) events for the same equipotential release distances. The Type 2-1 translocations are faster than Type 1 events and are shown in silver histograms which are also predominant occurrences for higher $d$. Type 2 events are relatively rare and the bar plots in the insets represent the occurrence of these three types of capture events on a percentage scale. }
\label{Capture1}
\end{figure}
\begin{table}[]
\begin{centering}
\begin{tabular}{c|cccc}\hline
~~$d$~~ & ~~~$\Phi$ ~~~&~~~~ $\mu$ ~~~~& ~~~~~$\lambda_{cap} = \frac{1}{\mu}$  ~~~~~& ~~~~~$\tilde{\lambda}_{cap}$ \\ \hline
4  & 0.322 & 1361.76 & $7.343 \times 10^{-4}$ & $7.618 \times 10^{-4}$   \\
8  & 0.157 & 2431.66 & $4.413 \times 10^{-4}$&  $4.112  \times 10^{-4}$   \\
12 & 0.101 & 3636.57 & $2.749 \times 10^{-4}$ & $2.858    \times 10^{-4}$   \\
16 & 0.071 & 5352.53 & $1.868 \times 10^{-4}$&  $1.897 \times 10^{-4}$    \\ \hline                                                   
\end{tabular}
\end{centering}
\caption{The table represents the fit parameters for the capture probability shown in Fig.~\ref{Capture-Prob}(a). $d$ is the vertical distances from the pore for the corresponding potentials $\Phi$. $\mu$ is the capture time distribution average of the released polymers from the different equipotential surfaces (Fig.~\ref{Capture-Prob}(c)-(f)) The fit parameters $\lambda_{cap}$ shown in the fourth column are used to fit the capture time graphs of Fig.~\ref{Capture-Prob}(a). The last column represent the best-fit parameters obtained numerically fitting $P_{cap}(\Phi, t)$.}
\label{capture_table}
\end{table}
\begin{figure*}[ht!]
\includegraphics[width=0.95\textwidth]{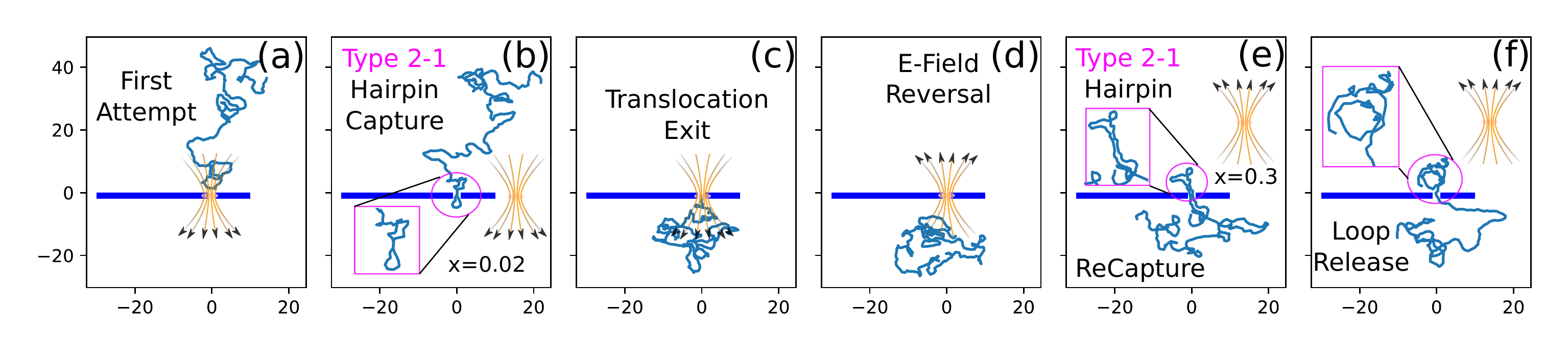} 
\caption{\small The series of snapshots show the multi-capture process of a single polymer by altering the voltage bias. (a) The first attempt of threading from the cis side (b) hairpin capture (Type 2-1), and (c) translocation from the cis side to the trans side. (d) After the reversal of the E-field (e) the same polymer gets captured from the bottom side of the pore. The hairpin loop structure is shown in the inset and at first both ends translocate at the same time. (f) Single end translocation begins after the unwinding of the loop.}
\label{Recapture-Config}
\end{figure*}
\par
Fig.~\ref{Capture-Prob} (a) also shows the dependence of the capture time on different equipotentials. A larger value of  the equipotential (closer to the pore) helps a faster capture. To get an improved statistics we monitor the journey of $1000$ independent fully flexible polymers after the initial release from randomly chosen coordinates on an equipotential $\Phi$ until they translocate. By increasing the release distance $d$ (lowering the strength of the equipotential $\Phi$), our study shows that there is a finite probability (shown in Fig.~\ref{Capture-Prob} (b)) that polymers drift away from the pore as the field strength becomes weaker at the distant equipotential surfaces. One can use this graph to define a capture radius~\cite{Slater2019}.
\par
Though a polymer is captured, its motion barely follows the curvature of a single field line going straight through the nanopore. A successful capture often requires multiple failed attempts but with an increasing number of failed attempts, the probability of capture also gets enhanced as the polymer gyration radius gets compressed by the E-field and remains in the vicinity of the nanopore opening.  Fig.~\ref{Capture-Prob}((c)-(f)) show the capture time distributions which sum up the polymer journey from release to capture including the failed attempts. With increasing $d$, the average capture time increases which suggests a longer wonder time when released from a weaker potential. The process of capture is Poissonian in nature and the distribution follows the shape of an exponential distribution where the mean and standard deviation is almost identical. It is worth mentioning that the exponential distributions of Fig.~\ref{Capture-Prob}((c)-(f)) produced by the BD simulation are the same as observed experimentally by Gershow and Golovchenko~\cite{Golovchenko2007}. 
\subsection{Modes of translocation} 
Dekker {\em et al.} introduced the nomenclatures for polymers threading into the pore with three different conformations (Type 1, Type 2-1, Type 2) events, depending on the relative location of the chain with respect to the pore~\cite{Dekker2005}. Mihovilovic et al. further studied and quantified translocation of these folded conformations~\cite{Stein2013}. We  observed these separate events in our simulation shown in Fig.~\ref{Capture1}((a)-(h)). In Type 1 events one end of the polymer threads (Fig.~\ref{Capture1}(a,e)), while in Type 2-1 a random location capture occurs. The symmetric threadings are Type 2 events (Fig.~\ref{Capture1}(b,f) \& (c,g)). Not only we observe these threading conformations in our simulation studies, in addition, our simulation provides the fraction of events belonging to these three categories shown  Fig.~\ref{Capture1}((m)-(p)) insets. For a closer release from $d=4.0~(\Phi=0.322)$, Type 1 capture percentage is marginally greater than Type 2-1 events but for all the other cases $d \geq 6.0$, Type 2-1 is the most abundant event. On the contrary, Type 2 events are rare occurrences ($\leq 2\%$) compared to the rest. These BD simulation results reveal  further details as well as validates earlier theoretical and studies experimental
studies~\cite{Muthukumar2018,Dekker2005,Golovchenko2007,Stein2013}.
\par

Recent theories~\cite{Muthukumar2018} indicate that translocation depends on the polymer's initial conformation and the degree of equilibration. The electric field and its gradient also influence the polymer shape during capture. We calculate the average end-to-end distance $\langle R_{end} \rangle$ of the polymer at the instance of capture to study its effect on the translocation time. During the pre-capture phase, a polymer undergoes stretching deformation due to the unidirectional field gradient and $\langle R_{end} \rangle$ deviates from the equilibrium end to end distance which is more prominent when starting from a distant equipotential from the pore mouth. The $\langle R_{end}\rangle $ histograms in Fig.~\ref{Capture1}((i)-(l)) become slightly right skewed for higher $d$ (lower $\Phi$) and $\langle R_{end} \rangle$ increases by $5-20~\%$ from the equilibrium average. However, the mean first passage time (MFPT) distributions of $1000$ independently captured polymers show a counter intuitive outcome, indicating a faster translocation time for higher $d$ where we previously observed that $\langle R_{end} \rangle$ is large. This apparent contradiction is resolved when we filter out Type 2-1 events from Type 1 events and plot the translocation time histograms separately for each type of event as shown in Fig.~\ref{Capture1}((e)-(h)). The translocation of Type 2-1 and Type 2 captured configurations are inherently different and faster than a single file translocation event as both ends of a hairpin loop configuration thread through the nanopore simultaneously until one end translocates completely and the loop unwinds. 
This phenomenon is also observed for the multi scan events as depicted in Fig.~\ref{Recapture-Config}(e) and in Section~F. For Type 1 translocation event, average MFPT and spread increase with $d$ (shown in the colored histograms in Fig.~\ref{Capture1}((m)-(p)), while for the Type 2-1 events a faster MFPT is obtained (silver histograms).
\subsection{Post-translocation compression}
Translocation being a faster process in presence of an electric field gradient, the polymer configuration gets compressed in the post translocation phage. Fig.~\ref{Rg-Ellipse} demonstrates the compression factor as the area of the gyration ellipse decreases more than 20\% compared to its pre-translocation stage. After  translocation, both the eccentricity and area of $\langle Rg \rangle $ ellipse remain constant but the fluctuation in $\langle Rg \rangle $ increases (pink cloud in Fig.~\ref{Rg-Ellipse}) as it enters into the diffusive domain.
\begin{figure}[ht!]
\includegraphics[width=0.48\textwidth]{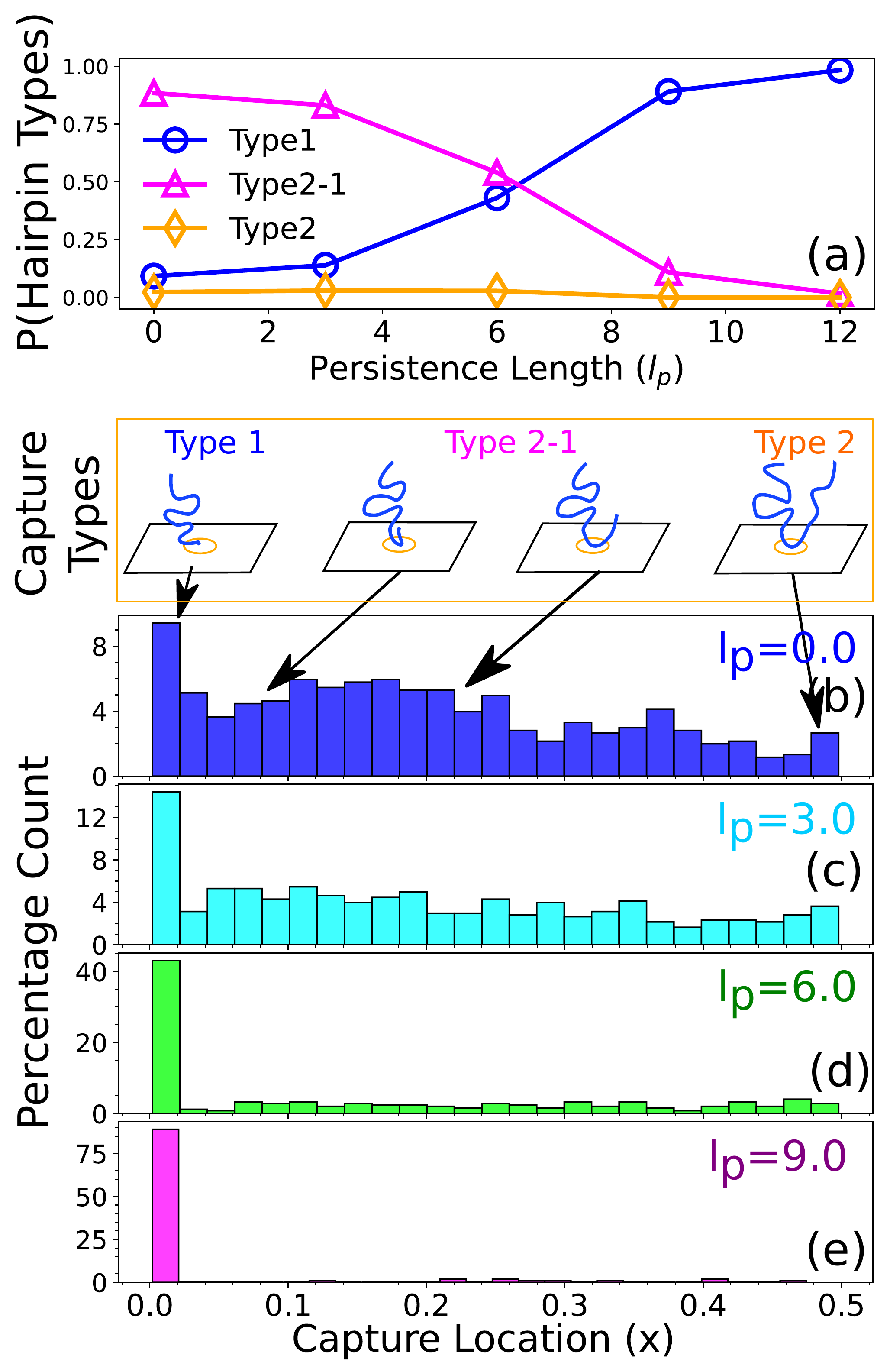}
\caption{\small (a) Hairpin capture probability for different stiffnesses of the polymer. By increasing the chain stiffness, the single file capture (Type 1 events) represented by the blue circles (\tikzcircle[blue, fill=none]{2.5pt}) increases while reducing the Type 2-1 events denoted by the magenta triangles (${\color{magenta}{\bigtriangleup}}$). The symmetric capture rate (Type 2 events) remains non-zero up to the chain persistence length $\ell_p=6$ and goes to zero for stiffer chains. (b)-(e) figures show the distributions of capture locations (in reduced units) for different persistence lengths. (b) For a fully flexible polymer, capture distribution is broader, Type 2-1 event occurrence has finite translocation probability along with the Type 1 events. The capture location distributions for stiffer chains are shown in (c) $\ell_p=3.0$ (d) $\ell_p=6.0$ (e) $\ell_p=9.0$ respectively. With the stiffening of the chain hairpin capture probability significantly reduces down improving the single file capture rate. } 
\label{Capture-Loc}
\end{figure}
\subsection{Multiple recaptures of a translocated polymer}It is the faster speed of the translocating DNA in a solitary nanopore that makes the current blockade measurements noisy for sequencing purposes. To overcome this issue, multiple recaptures~\cite{Golovchenko2007} of the same molecule can be a viable option that relies on increased statistics, hence, enhancing the accuracy of the measurement. In our simulation setup, we reverse the voltage bias after a successful translocation as the center of mass of the polymer moves $20 \sigma$ away from the pore  and study the polymer dynamics as a function of its persistence length.  During the multi-capture events statistics are collected for six independent runs, each containing $100$ scans unless the polymer drifts away. Fig.~\ref{Recapture-Config} demonstrates the recapture events for a polymer having the persistence length $\ell_p=3.0$. Our study shows that even in the extreme narrow nanopore limit (pore diameter of $2 \sigma$), hairpin capture probability dominates over the single file translocation events for semi-flexible polymers (see the simulation movies). In addition, from Fig.~\ref{Capture-Loc}(e) it is evident that only beyond $\ell_p=6.0$, Type 1 capture probability is higher than hairpin capture (Type 2-1 and Type 2) probability and $\ell_p=6.0$ serves as a critical point between these two events.
To understand how the chain stiffness affects the capture process in presence of a field gradient we have studied  the capture location distributions shown in detail in  Fig.~\ref{Capture-Loc}(a)-(d). For a fully flexible chain ($\ell_p=0$) all three types of capture occur but with the increasing stiffness, as expected, the single file captures become predominant. It is important to note that for $\ell_p=3.0$; Fig.~\ref{Capture-Loc}(c) - that corresponds to the persistence length of a dsDNA under most experimental conditions the distributions of the capture locations obtained from the BD simulation closely resemble those obtained experimentally by Mihovilovic {\em et.al.}~\cite{Stein2013}.
\par
We further verify that for a longer chain with $N=512$, the capture location distributions for different stiffness are qualitatively similar, however, the probability of the Type 1 capture increases compared to the other two types of events. This enhancement of Type 1 events with the increased chain length is more prominent for the fully flexible polymer and does not alter the distribution as such for the cases with higher persistence lengths, where Type 1 is the only predominant event. 
\subsection{Enhancement of capture rate by End Tagging}
\begin{figure}[ht!]
\includegraphics[width=0.48\textwidth]{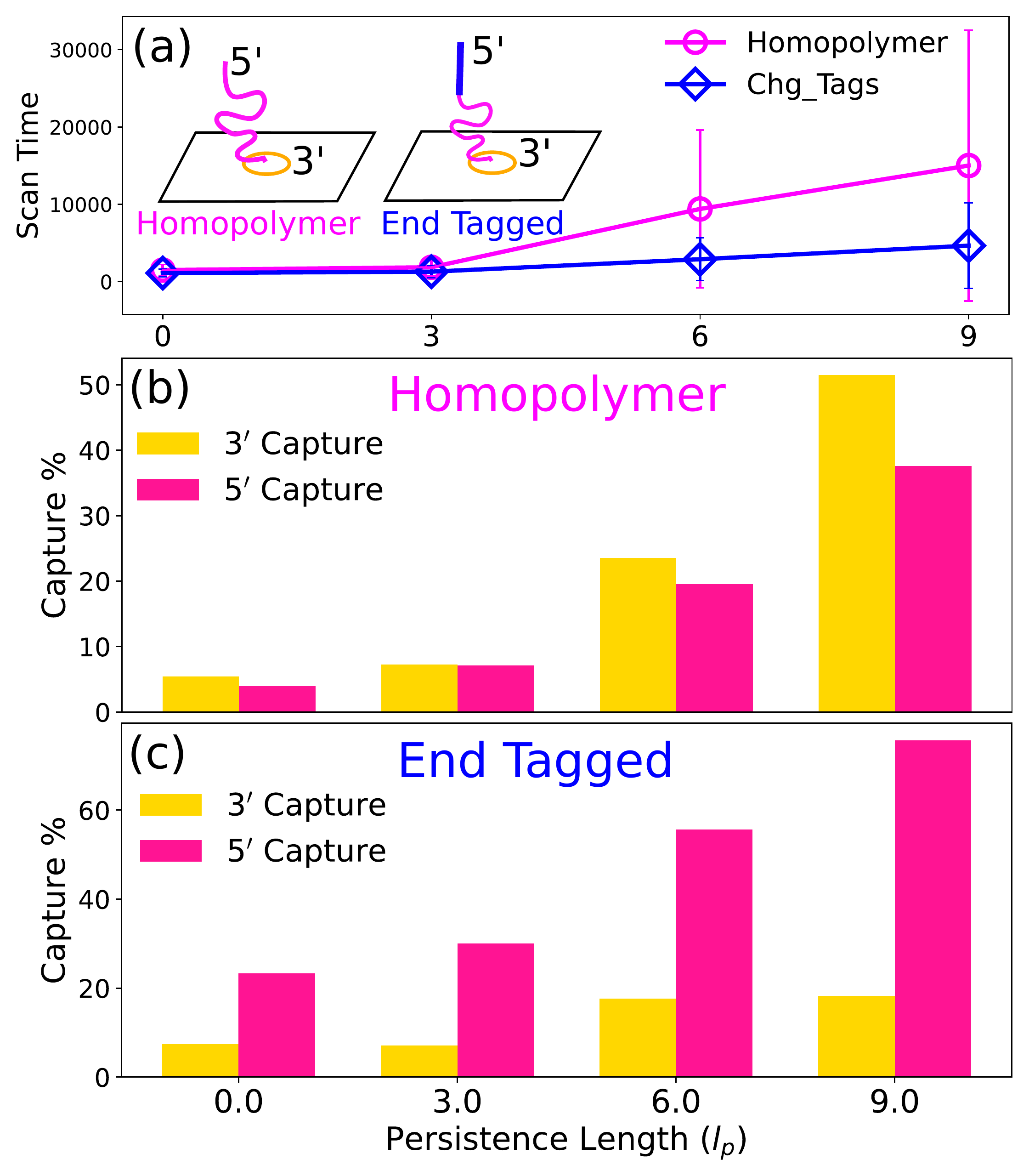}
 \caption{\small  (a) Figure shows the scan time duration of a homopolymer in magenta circles (\tikzcircle[magenta, fill=none]{2.5pt}) and polymer with end tags in   blue diamonds (${\color{blue}{\Diamond}}$) as a function of the polymer persistence length. (b) Homopolymers of different stiffnesses are captured and scanned multiple times and the capture percentage of $3^{\prime}$ and $5^{\prime}$ end terminus are shown in yellow and pink bars. (c) The same bar diagrams are depicted for polymer with charged tags located at the $5^{\prime}$ end.} 
\label{End-Tag}
\end{figure}

~Attaching a ``charged motif'' to a dsDNA provides additional information about its translocation dynamics through a nanopore~\cite{Rosenbloom2004, Reisner2019}. In our simulation, we enhance the charge content of the last six monomers of one end (let's call it $5^{\prime}$ end) by three times of a normal monomer ($q_{tag}=3q$) and apply the same recapture method to demonstrate the effect of end tagging on the capture and translocation process. We first monitor the scan time duration (capture time + translocation time) of a homopolymer varying its persistence length. Fig.~\ref{End-Tag}(a) (magenta circles) show that the scan time increases non-linearly with an increase in the chain stiffness and the error bar in scan time widens which signifies a large 
variation in capture time for the stiffer chains. To reduce the uncertainties of capture, we use the end tagging method (see the insets of Fig.~\ref{End-Tag}) and blue diamonds confirm that end tags not only conclusively lower the scan time for the stiffer chains but the error range also gets reduced. 
\par
In reality, without the data post-processing, it is almost impossible to know which end of the polymer threads into the pore during each recapture scan. By charge tagging the $5^{\prime}$ end, we potentially break the degeneracy as the tags with high charge content are prone to enter into the nanopore first due to the stronger pull of the electric field. Fig.~\ref{End-Tag}(b) shows that the capture probability of the $3^{\prime}$ and $5^{\prime}$ ends are almost equal for a homopolymer even with different stiffnesses, hence degenerate. Our simulation confirms that these degeneracies are broken and $5^{\prime}$ end has a significant higher probability of entrance when tags are present at the $5^{\prime}$ end as shown in Fig.~\ref{End-Tag}(c). This tagging prioritizes the $5^{\prime} \rightarrow 3^{\prime}$ uni-directional translocations from either side of the pore during multi-recapture and scan which can potentially benefit the single nanopore barcoding and sequencing experiments where uni-directional reads would be preferred. Moreover, one can make the end terminus oppositely charged which would in principle further promote the uni-directional translocations due to the favorable attractive interaction at one end and a repulsive force on the other end in presence of an extended electric field.
\par
\section{Discussion and conclusion} In this article, we report the details of the capture and translocation of a DNA polymer as it meanders its way to the pore entrance from different equipotentials. Capture from different equipotentials affects the conformations at the pore entry as well as the translocation speed. Our studies reveal the details of multiple failed attempts and eventual success in presence of an extended E-field, quantify and differentiate  accurately characteristics of translocation of both straight and folded conformations. We analyze further details of the folded coordinates in reference to the entire chain. Based on these results, we suggest how one can take advantage of it to improve the accuracy of the experimental protocol using a solitary nanopore. We also demonstrate the model is capable of reproducing all the details of an actual experiment~\cite{Golovchenko2007} and Mihovilovic {\em et al.}~\cite{Stein2013}. We introduce a charge tag at one end of the dsDNA and demonstrate that the scan time and its variability are significantly reduced especially for the stiffer chains and uni-directional capture probability increases by lifting the degeneracy of which end threads through the nanopore during the multi-capture process. We expect this work will promote future experimental studies on repeated scanning of a biopolymer through nanopore and can be useful for a large community involved in various types of biopolymer translocations.
\par
\section{Acknowledgment}
The research has been supported by grant number 1R21HG011236-01 from the National Human Genome Research Institute at the National Institute of Health. All computations were carried out using STOKES High-Performance Computing Cluster at UCF. The simulation movies are generated using the Visual Molecular Dynamics package~\cite{VMD}.

\vfill
\end{document}